# Dissecting a Social Botnet: Growth, Content and Influence in Twitter

**Norah Abokhodair**
The Information School
University of Washington
Seattle, WA, USA
noraha@uw.edu

**Daisy Yoo**
The Information School
University of Washington
Seattle, WA, USA
dyoo@uw.edu

**David W. McDonald**
Human Centered Design & Engineering
University of Washington
Seattle, WA, USA
dwmc@uw.edu

**ABSTRACT**
Social botnets have become an important phenomenon on social media. There are many ways in which social bots can disrupt or influence online discourse, such as, spam hashtags, scam twitter users, and astroturfing. In this paper we considered one specific social botnet in Twitter to understand how it grows over time, how the content of tweets by the social botnet differ from regular users in the same dataset, and lastly, how the social botnet may have influenced the relevant discussions. Our analysis is based on a qualitative coding for approximately 3000 tweets in Arabic and English from the Syrian social bot that was active for 35 weeks on Twitter before it was shutdown. We find that the growth, behavior and content of this particular botnet did not specifically align with common conceptions of botnets. Further we identify interesting aspects of the botnet that distinguish it from regular users.

**ACM Classification Keywords**
H5.m. Information interfaces and presentation (e.g., HCI): Miscellaneous.

**General Terms**
Human Factors.

**INTRODUCTION**
During the past two years, Laila has been interested in the Syrian uprising. Laila is studying political science and her Arabic is a little rusty, but good enough to search and read tweets. Over time she has observed users who tweet in support of the Syrian government and others who tweet in favor of the Free Syrian Army (FSA). Laila has noticed that users who tweet from a given perspective will share common followers and will retweet each other.

Today, one particular tweet catches Laila's eye. The tweet looks like a news headline favorable to the Syrian government. Laila clicks a shortened URL in the tweet and finds herself on a news website that is known to be backed by the Syrian government. Laila suspects this information might be biased and returns to Twitter to search again. In the results of her search, she notices that the tweet has now been retweeted thousands of times. Laila thinks to herself, "Who are these people?" and wonders why this tweet is so special. She traces the topic to uncover the originator of the tweet. The profile for the account is sparse, with few details about the user.

Online social spaces often challenge users' abilities to understand others, interpret their actions and collaborate. This challenge is not particularly new. However, what is a new and growing challenge is that we can no longer assume that a social space is populated exclusively by people. Bots have become significant players in online social spaces. A recent study estimated that 61.5% of total web traffic comes from bots. One recent study of Twitter revealed that bots make for 32% of the Twitter posts generated by the most active account [5]. In the future, more and more of the social interpretations that we will be challenged to make will be about collections of people interleaved with collections of machines acting like people. In the context of CSCW it is essential for designers and researchers to anticipate and design for the interactions among bots and human participants to improve the potential collaborations.

This study considers one specific social botnet in Twitter to understand how it grows over time, how the content of tweets by the social botnet differ from regular users in the same dataset, and lastly, how the social botnet may have influenced the relevant discussions. Our botnet data comes from a much larger corpus of tweets focused on the Syrian civil war. The botnet emerged from an unrelated set of social network explorations where we were considering the structural properties of tweeting and retweeting around specific events of the Syrian civil war. Our analysis follows the botnet comprised of 130 user accounts, from the earliest evidence we have of its existence through the 35 weeks it was active, up to the day that Twitter suspended the botnet.

In the following we first define a social botnet and cover the prior research related to studies of social bots and botnets. Next, we describe the dataset of tweets in which the Syrian social botnet (SSB) was discovered. We then describe three analyses we conducted on the data set; (i) growth of the Syrian social botnet over time, (ii) a grounded content analysis of botnet and regular user tweets, and (iii) the potential influence that the botnet had within the context of the data set.

**BOTS AS SOCIAL ACTORS**

A bot is software designed to automate a task in a computing system. The popularity of bots has risen over the last few years with the rise of web services that simplify programmatic interactions with a wider range of web based resources and a growing corpus of code libraries that implement the onerous, repetitive aspects of accessing a web service.

We characterize a social bot as an *automated social actor* (ASA). This is software designed to act in ways that are similar to how a person might act in the social space. The level of sophistication and the roles that a given ASA will fill in the social space can vary. An ASA could be relatively simple, like bots that aggregate information from web news and re-present it as a set of tweets in Twitter, or quite sophisticated like a conversational bot that is attempting to pass a Turing test. An ASA can be malicious by trying to disrupt a social space, or the ASA can try to augment and support a social space, say by explaining different cultural norms. An ASA becomes more interesting when it attempts to presents itself as an actual person in a social space. That is, when ASAs attempt to pass as an actual person, goals, motivations, and mechanisms behind the bot are interesting to uncover, unpack and understand. One danger of ASAs is passing as a person or organization to promote specific ideologies to create a false sense of consensus.

Our definition draws a subtle distinction between ASAs and the way "botnet" has been predominantly used in the popular press. The popular use of botnet describes a set of physical computers, machines, that have been compromised by malicious software, a virus, that allows remote control of the machine. Often these botnets are used for spamming, phishing, distributed denial of service attacks (DDoS) or other nefarious activities. A relatively recent survey article on botnets covers the research literature that focuses largely on these cyber-security aspects of bots and botnets [14], but also contains some valuable nuggets related to our focus on bots as social actors.

Our definition is closer to the conceptualization of early Internet bots as social entities that would populate IRC (internet relay chat), or MUDs (multi-user domains). Our definition can cover the current botnets involved in astroturfing[1], as well as the range of social and informational uses. We take analytical stance that botnet activities could be deemed good or bad, depending on the social context and the types of actions that the bots take. In our view, ASAs are merely another part of the online social space with which we, as humans, must understand and interact.

In the context of CSCW, understanding bots and the way bots work will directly influence the way we think about the design of collaboration systems and how we analyze the subsequent social activities that occur within those systems. As bots become part of collaborative systems, an effective analysis should be able to differentiate "real" human participation from that of bots, necessitating more sophisticated means of detecting social and collaborative activities of bots and bot networks.

**STUDYING SOCIAL BOTS AND BOTNETS**

Bots themselves are not a particularly new phenomena. In fact, some of the earliest bots and botnets were social bots; ASAs. In the following section we review related literature on social bots. We structure our review around the social infrastructures in which bots have participated; Internet Relay Chat (IRC), Multi-User Domains (MUDs), Massive Multiplayer Online Games (MMOGs) and ending with contemporary social computing systems.

**Bots on Internet Relay Chat (IRC)**

Far before ASAs began invading our current social network systems like Facebook or Twitter, they populated IRC. One of the earliest known IRC bots, Eggdrop, emerged as early as 1993 [16]. This bot could be configured to provide a range of triggered actions that helped administer IRC channels. The slightly social aspects of the bot include welcoming and greeting new participants and warnings for some user actions. There have been many bots developed for IRC, with ever increasing social sophistication. One reason that bots were traditionally popular on IRC was due to the simplicity of implementation and ability of scaling IRC [11].

A majority of the studies on IRC bots and botnets have focused on detection techniques, tracking and characterizing botnets. Abu Rajab et al. [1] used a multifaceted approach to track and capture the behavior of IRC based botnets. The study revealed a number of behavioral and structural features of IRC botnets, for example, the authors identified different scanning mechanisms (e.g., uniform, non-uniform, and localized scanning algorithms) botnets used to target its victims.

---

1. Astroturfing is a fake "grassroots", populist, campaign in a social media space. This is named after a type of fake field turf (fake grass) that was created in the 1960s.

In particular, the authors tracked 192 unique IRC botnets during the three months measurement period and concluded that botnets are generally "long-lived" in reference to the average lifetime of 47 days. In our study, we were able to capture a live botnet and actively track its behavior over more than six months until it was shut down, which adds an interesting case study of an exceptionally long-lived botnet. On botnet size, the authors conclude that it is highly likely that the botnet footprint is greater than the node size, which confirms the need to understand and follow this phenomenon on social networks to reduce potential negative impacts and raise awareness about its implications for the social graph.

**Mobs in Multi-User Domains (MUDs & MMOGs)**
The advent of Multi-User Dungeons (also Domains) (MUDs) included the need for automated social agents (ASAs) to enrich the playing experience. These mobs (mobiles) were not simply passive monsters that players could kill for experience points. Mobs were towns' people, shopkeepers, bards, and government officials who had their own (programmed) motives and goals. Players could learn about the environment and progress the game through interaction with mobs. But as well, mobs could play key roles regulating the social environment of the game [2,11]. In World of Warcraft (WoW) mobs often shared state and comprised a very simple botnet. Despite solid studies on the social aspects of gaming, little work has focused on the specific social roles played by the mobs (bots) themselves.

With the rapid growth of online gaming market (e.g., MMOGs such as WoW) came an increase in the use of unauthorized game bots [9,19]. These bots provided enhancements and triggering mechanisms for individual players often sitting in between the players' client application and the game server. Some bots could effectively play the game completely in the players' absence. These unauthorized bots were often considered a form of illegitimate play sabotaging game ecologies (i.e., amassing game currency, or experience points).

Yamlolskiy and Govindaraju [19] articulate a typology of these types of game bots reflecting various genres of online games. In this MUD/MMOG space they articulate the idea of a botnet as a collection of bots which share information and coordinate to provide players who employ those bots a competitive advantage.

The concept of federation or collusion among ASAs is important to our study. We leverage this botnet concept to understand how the ASAs in our study work together in their attempt to influence the direction and trending topics in an online social space.

**Bots in Contemporary Social Computing Systems**
The emergence of social botnets on social network systems (SNS) sparked interest amongst researchers to study this phenomenon and understand it is implications for the social sphere. In 2012, a report by Facebook estimated that 5-6% of all Facebook accounts are fake or bogus, which is considered a high percentage for Facebook as it means that approximately 50 million users are fake [14]. One study by Boshmaf et al. [3] discussed the use, impact, and implication of social bots on Facebook. In their research the authors designed and built a Socialbot Network (SbN) on Facebook to understand the ways in which social bots could threaten SNS by large-scale infiltration. The research focused on social bots that mimic human behavioral patterns to expose private information and destroy trust amongst users rather than bots that raise concerns through spamming. This research showed how vulnerable Facebook is to sophisticated social bots and how they can spread without being flagged by Facebook users and marked by the Facebook Immune System, which is a real-time system embedded in Facebook to protect its users and social graph from malicious spam [18].

The Web Ecology Project [19], called on researchers to participate in a competition to design social bots that have constructive impact on social networks to explore the different ways in which a social bot could influence online discourse. The results of this competition showed that social bots were able to reshape the social graph on Twitter by influencing and initiating some conversations with accounts that were not following each other previously.

Other studies have considered the social implications of bots for the platforms where the bots interact. Geiger [8] elaborates the social roles surrounding bots in Wikipedia, but does not consider how bots interact socially with users within Wikipedia.

While the research projects above explored the potential use cases of SNS based social botnets in general (in both positive and negative manners), we are particularly interested in the political use of social botnets. Recently, following the Arab Spring in 2010, SNS emerged as a political platform for activism. For example the dissemination of news, images and real-time videos from the streets of Egypt in 2011. The simplicity and receptivity of Twitter allows it to be used in many ways creatively by anyone online with good or bad motives. In this regard, a study by Ratkiewicz et al. [15] investigated how Twitter can be exploited through astroturfing campaigns. In their study the authors' analyzed Twitter data that they obtained from an astroturfing detection tool named "Truthy". They concluded that it is vital to detect the astroturfing campaigns before they go viral on Twitter because once they pass the first stages of detection by Twitter's spam detector it is easy to abuse and spread misinformation that can result in a false sense of broad agreement.

**THE SYRIAN SOCIAL BOTNET (SSB)**
The related work suggests that social botnets have far-reaching implications and they are only getting more sophisticated in terms of scanning for victims and mimicking human behaviors. Much work has focused on automatic detection mechanisms [3, 15] but little has been done to understand the behaviors and characteristics of social botnets in social computing platforms. This paper aims to elaborate our understanding of the latter.

We captured a Twitter based botnet, in the wild, based on hashtags related to the Syrian civil war in April 2012, which we call the Syrian social botnet (SSB). We tracked the SSB for more than six months during the ongoing civil war, from April to December 2012. In this paper we focus on the measurement of influence, structure and growth of the SSB. Additionally, we present the results of our content analysis on a collection of 3000+ English and Arabic tweet samples. To the best of our knowledge this is the first such study on this topic.

**Method: Discovering a Social Botnet**
We initially started a collection of tweets related to the Syrian civil war in April 2012, approximately one year after the earliest peaceful protests related to the war. We searched for terms related to Syria (the country) and the names of five major Syrian cities (Damascus, Aleppo, Idlib, Hama, Homs) using both the English and Arabic names for each city as well as the country. This initial dataset covers 18 months with between 150K and 500K tweets per day, and approximately equal proportions of English and Arabic tweets. We relied on retweet data to begin a social network analysis because that action is more indicative of attention to content of the tweet [4]. During that analysis we focused on the social structure of specific events during the Syrian civil war (e.g., Houla massacre on May 25, 2012, anniversary of 9/11 attack, Damascus massacre on December 12, 2012). From those events we noticed recurring clusters of users retweeting each other (for an example see Figure 1). Those recurring clusters uncovered the botnet in this study. We examined clustering patterns and identified a set of 20 high-volume twitterers and their social connections based on retweet activity. Next, we traced the profile information and activities of each high-volume Twitterer. We found that 17 accounts had been suspended and had stopped tweeting altogether on the same date and time; around 6:30 AM UTC, November 20, 2012. We examined the content of their last tweets and found that all of the suspended accounts were retweeting a unique news aggregator bot: @GB1.[2] We conducted a preliminary analysis on the social structure around @GB1 and 17 suspended user accounts. Specifically, we snowballed the retweet connections backwards in time to identify 130 usernames, comprising a social botnet, which we named the Syrian social botnet (SSB).

This study focuses principally on the 35 weeks that the botnet was active. We conducted a comprehensive analysis aimed at understanding (i) the growth and structure of the botnet over time, (ii) how the content of the botnet (in Arabic) differs from the content of regular users tweeting in Arabic or English, and lastly, (iii) how the botnet may have influenced the overall discussion.

**Growth and Structure of SSB**
We wanted to understand how the botnet grew and whether 130 bots all played the same role or whether they behaved differently over time. For each identified bot, we examined (a) the total number of tweets per week, (b) the total number of retweets per week, and (c) the total number of "bot-retweets" per week, which is the portion of retweets of other SSB bots. Upon examining these features, we graphed and analyzed the trend of bot behaviors over the 35 week period (see Figure 2). Each bot was classified into one of five categories based on distinct behavioral patterns: short-lived bots, long-lived bots, generator bots, core bots and peripheral bots.

*Core bots* are the Twitter user accounts that we consider as primary agents of the SSB (see Figure 2-D). They typically demonstrate the following behavioral patterns: First, core bots tweeted frequently. They normally generated more than 1600 tweets per week (M=1641), which is roughly a tweet every 6 minutes nonstop. Moreover, core bots accelerated their rates of tweeting as the weeks progressed. In particular, on week 28, the median rate of core bot's tweeting reached 5733, which is roughly a tweet every 1.8 minutes (see Figure 1 and Figure 2). Second, roughly half of a core bot's activity focused on retweeting (the ratio of bot-retweets to total tweets: M=0.4691). Particularly, a core bot would only retweet which was created by its fellow core bots in the SSB (the ratio of bot-retweets to total retweets: M=0.9909).

During the 35 week period, the lifespan of core bots varied from 1 to 32 weeks (M=10). In particular, we identified 23 core bots that survived more than 25 weeks (M=28), and we marked them as *long-lived bots* (see

---

2. For Twitter data reporting, we are using a pseudonym to reduce searchability and enhance confidentiality.

Figure 2-C). Conversely, we identified 43 *short-lived bots* that survived less than 6 weeks (M=2) during the first 11 week period (see Figure 2-A). Long-lived bots and short-lived bots are sub types of core bots.

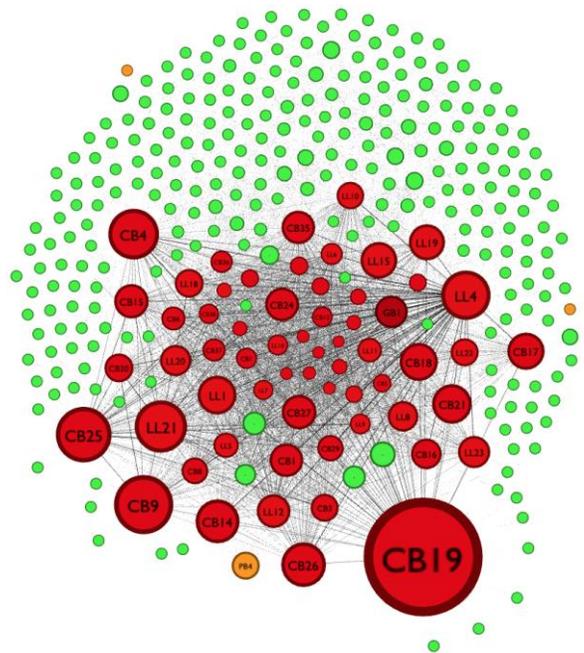

Figure 1. Social network of week 28 (red=core bots and long lived bots; orange=peripheral bots; green=other twitterers).

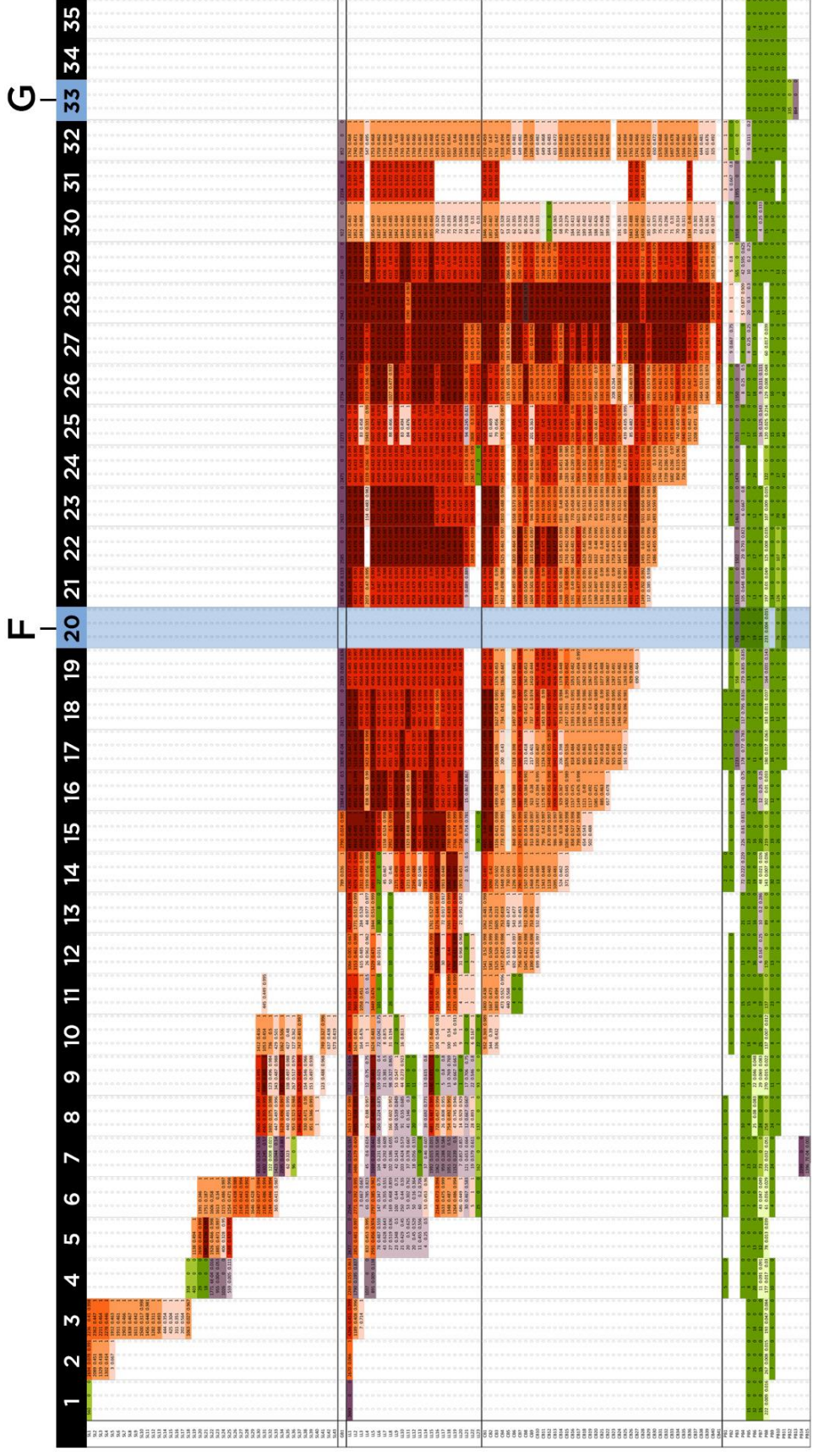

Figure 2. Behavioral pattern analysis of the SSB. We identified five distinct types of the SSB bots: short-lived bot (A), generator bot (B), long-lived bot (C), typical core bot (D), and peripheral bot (E).

*Peripheral bots*, on the other hand, are potentially legitimate twitterers that we assume to be unwittingly complicit in the SSB by retweeting one or more tweets generated by the core bots. Legitimate users being lured and participating in the network is often observed in astroturfing [15], therefore, we expected that this would also happen in the SSB. Since their pattern of behavior is distinctly different from the core bots and since we cannot unequivocally verify that each of those accounts belong to a "real person," we will call them peripheral bots as opposed to assuming they are actual human users (see Figure 2-E). In contrast to core bots, peripheral bots tweeted occasionally, normally generating no more than 70 tweets per week (M=11). In addition, the ratio of bot-retweets to total retweets was significantly lower for peripheral (M=0) compared to core bots. In other words, peripheral bots seldom retweeted the other bots in the SSB. While core bots temporarily stopped tweeting altogether on week 20 peripheral bots continued to tweet during that void (see Figure 2-F). We identified 15 peripheral bots out of 130 bots total in the SSB.

In addition, we identified one exceptional type of core bot, a *generator bot* (see Figure 2-B). The generator bot typically produced more than 2100 original tweets per week (M=2384.5, that is, constantly posting more than one tweet every 4 minutes). As opposed to typical core bots, the generator bot seldom retweeted anything (the ratio of bot-retweets to total tweets: M=0). However, this bot (@GB1) was heavily retweeted by the other core bots. Despite these atypical traits, we included the generator bot as a subset of core bots based on (a) its high tweeting frequency, and (b) its behavior on ceasing to tweet during the week 20 and after the week 33 shutdown (see Figure 2-G).

Overall, we observed continuous growth of the SSB over the 35 week period. On week 1, we observed only two active core bots. By week 28, however, the number increased to 64 core bots, and the SSB emerged as a predominant cluster shown on the general Syrian retweet network graph (see Figure 1). The SSB's networking activity (including numbers of total tweets, retweets, and bot-retweets) gradually increased over time as well. In particular, the volume of SSB's networking activity spiked on week 28. However, despite this immense volume, the SSB was not immediately shutdown. On week 33, which is five weeks after its peak, the SSB was suspended by Twitter. Unsurprisingly, a close parallel to this growing pattern is found in the SSB's influence on general Twitter community (for details, see the Influence of SSB section).

**SSB Content Categories**

Another key question for understanding the SSB is how the content of the botnet differs from regular users tweeting about the Syrian civil war. We conducted a grounded content analysis of a random sample of tweets from English twitterers, Arabic twitterers, and the botnet (which tweets in Arabic) and compared the difference between these three types of tweets.

We created a set of grounded categories in several rounds. In the first round we selected a little over 100 tweets at random from the botnet. One coder (a native Arabic speaker), read through the tweets and proposed a draft coding manual. After that the research team met to discuss the codes, leading to a first revision of the initial set of codes. At the outset, two overarching coding categories were identified: Primary Code based on topic and Secondary Code based on geographical context. Initially, we started with a small number of primary codes and two context codes (local and international). The categories and tweet exemplars were then reviewed and discussed with native English speakers to clarify translational, colloquial, and structural issues in the proposed categories to make sure they could fit with English tweets. In a second round, we took a random selection of English and Arabic tweets to test and elaborate the categories from the first round.

*Primary codes*

Our efforts resulted in 13 primary code categories as follows:

*1.1. Opinion.* Self-expressive remarks about the ongoing Syrian crisis, including (but not limited to): opinions typically based on personal analysis of the situation, patriotism, critique or criticism, rhetorical questions, jokes that people use to self-express and de-stress from the crisis, prayers for the people of Syria, and poems or similar attempts at artful expression. If the text of the tweet is an excerpt from a news article (i.e., that you can tell by viewing a link and reading the story), then the tweet should be coded "News".

*1.2. Testimonial.* Testimonial remarks based on personal experience of the Syrian civil war, including references to one's family members, close friends and acquaintances. May link to amateur videos shot during the war (e.g., footage of a battle, funeral or aftermath). If a video link contains any institutional symbols such as a unit name or unit logo, that suggests persistent presentation of information relative to that unit, code as follows: (1) If the video conveys immediacy ("happening now") then code as "Breaking News" instead of Testimonial; (2) If there is no immediacy (i.e., a montage of recorded events) then code as "News" instead of Testimonial.

*1.3. Conversation.* Conversational remarks that are clearly part of a dialog between two or more people, which cannot otherwise be coded without having access to the whole conversation. One strong indicator is when a @username is first in the post. To test if a tweet falls in the Conversation category, remove the @username and judge whether the tweet could be coded into any other category. If without the @username it could fall into one of the existing categories, then the tweet should be coded as that category. If without the @username the tweet would be "Uncodeable" then code the tweet as "Conversation".

*1.4. Breaking News.* Breaking news, information, on the Syrian civil war, typically uses emphasized headings such as "URGENT", "BREAKING NEWS", "IMPORTANT", and "NOTE" to indicate immediacy. Any tweet that is generated by a news aggregator type of user account, where the user account is labeled with some form of "breaking news" is coded here. Headings such as "emergency", "help" or "please help", are not coded here; these terms likely suggest "Mobilization of Resistance/Support" category.

*1.5. News.* Tweets relating information about the Syrian civil war from an "objective", third party, stance or that are clearly trying to carry an objective tone. Includes tweets from verified news accounts and news aggregator accounts. This category includes reports from NGOs and martyr (obituary) notices. If there is a link in the tweet and that link points to a news article from a reliable source which is either "analysis" of news or "editorial" about news, then the tweet is coded in this category.

*1.6. Mobilization of Resistance/Support.* Tweets that make calls for organizing, meeting, protesting and gatherings related to the civil war. This includes online petitioning, retweet requests and related forms of slactivism.

*1.7. Mobilization for Assistance.* Tweets mobilizing people and organizations for social and humanitarian aid such as donations of food, clothing, and money.

| Content Categories | Botnet | Arabic | English |
|---|---|---|---|
| News | 538 (52.6%) | 359 (35.0%) | 376 (37.6%) |
| Other | 325 (31.8%) | 29 (2.8%) | 39 (3.9%) |
| Opinion | 127 (12.4%) | 465 (45.3%) | 258 (25.8%) |
| Spam/Phishing | 13 (1.3%) | 10 (1.0%) | 249 (24.9%) |
| Testimonial | 0 (0.0%) | 54 (5.3%) | 14 (1.4%) |
| Conversation | 1 (0.1%) | 23 (2.2%) | 20 (2.0%) |
| Breaking News | 4 (0.4%) | 47 (4.6%) | 18 (1.8%) |
| Mobilization of Resistance/Support | 3 (0.3%) | 9 (0.9%) | 7 (0.7%) |
| Mobilization for Assistance | 1 (0.1%) | 4 (0.4%) | 2 (0.2%) |
| Solicitation for Information | 1 (0.1%) | 4 (0.4%) | 1 (0.1%) |
| Information Provisioning | 2 (0.2%) | 19 (1.9%) | 12 (1.2%) |
| Pop Culture | 7 (0.7%) | 3 (0.3%) | 4 (0.4%) |
| Total Tweets Coded | 1022 | 1026 | 1000 |
| Uncodable (removed from analysis) | 26 | 11 | 207 |

**Table 1. Comparison of tweet content categories between SSB, regular Arabic and English twitter users.**

*1.8. Solicitation for Information*. Tweets asking about the situation in Syria. Rhetorical questions about the civil war are coded as "Opinion".

*1.9. Information Provisioning*. Tweets that provide information that is individually actionable.

*1.10. Pop Culture.* Tweets that reference celebrities, music, sports and entertainment and related people.

*1.11. Other*. Tweets related to Syria and other countries, but not clearly associated with the civil war.

*1.12. Spam/Phishing*. Tweets that look to be some form of spam or phishing.

*1.13. Uncodeable*. Tweets in languages other than Arabic or English or too short to reasonably interpret and code in another category.

*Secondary codes*
Additionally, we developed the two code categories based on context. These codes are independent from the primary codes:

*2.1. Local Context.* Tweets that mention the current situation in (and only in) Syria that is otherwise unspecified. Including: key events in Syria, elections, and military reaction.

*2.2. International Context.* Tweets with any mentions of foreign countries surrounding Syria and international organizations such as UN and Amnesty International. This category includes mainly: reactions from other countries to the situation in Syria, mobilization in support for Syria in other countries and hashtags listing many foreign countries.

These categories were developed based on an iterative exploration of samples of tweets, assumptions about the way a botnet might work, and what we have observed in the prior literature. One assumption we had was that the botnet would attract attention through use of popular culture references, resulting in the Pop Culture category. Categories like our Information Provisioning category have been reported in crisis informatics research as a way that social media users have of swiftly finding rescue or aid [17]. As well, the Mobilization for Assistance and Mobilization of Resistance/Support categories are similar to content categories from research findings on the role that Internet-based technologies play within social mobilization and political processes [6]. In the case of Opinion and Testimonial categories, we noticed a difference between people commenting about other media, such as a photo, a video or a news article, and video footage or images that were most likely recorded by the twitterer himself. By distinguishing between first-person content and comments on others' content we sought to offer a more nuanced analysis of differences in the botnet, Arabic and English content around this same event.

We used a Cohen's kappa to help measure our coding consistency over primary codes, which are independent from the secondary codes. Cohen's kappa is one measure of inter-coder reliability. Values range from 1.00 to -1.00, with 1.00 indicating perfect agreement and 0.00 indicating agreement that is no better than chance [7,12]. Two commonly referenced benchmarks for interpreting the values of Cohen's kappa are Fleiss, Levin & Paik [7] and Landis & Koch [12]. Cohen's kappa was calculated for the two coders based on a random sample that of 120 English tweets. Our sample was generated from English tweets only due to difficulty in finding a second fluent Arabic speaker. The overall score was k=0.64, indicating "intermediate to good" or "substantial" agreement by the respective benchmarks. We had better agreement in a few categories (News, k=0.74; Spam/Phishing, k=0.75) and poorer agreement in a few categories (Other, k=0.49; Opinion, k=0.50). But overall, the agreement between the two coders for this qualitative coding was relatively high.

### Content Analysis: SSB vs. English and Arabic

The coding process considered the content of the tweet, the content at the end of any link or URL in the tweet, and the username of the account as appropriate. In cases where the content at the end of a link had been removed, taken down or made private and unviewable, then the coding decision was based on the content and username alone.

Next, we generated three random samples of tweets; a little over 1000 tweets each from Arabic, English and SSB. Retweets were excluded from these three samples. The random samples were generated in multiple rounds of between 100-200 tweets each. The goal was to have at least 1000 coded tweets from each of the three categories, excluding any tweets from the total which fell into the "Uncodeable" category. A native Arabic speaker coded both the SSB and Arabic tweet samples. The English tweets were coded by a different coder. The team discussed difficult and troublesome tweets in each round to come to a consensus code for the most difficult tweets.

Table 1 illustrates a number of interesting content differences between the tweets by the SSB, regular Arabic twitter users and regular English twitter users. Table 1 also includes the number of tweets that we coded as Uncodeable. This number was quite high in the English sample because in that set there are many tweets that use English terms or hashtags that are names of Syrian cities, but which are actually in another foreign language. For instance, the term "hama" stands out significantly as a Latin character term that often accompanies tweets in Japanese.

One should immediately notice that a little more than half of the tweets by the SSB are News, whereas for regular users (both Arabic and English) News only comprised a little over one third of the total. It is useful to keep in mind that this analysis does not include retweets, so these tweets are often some representations of a news story or an attempt to report or describe events in an objective tone. In the case of the SSB, we noticed that many News tweets led to one of three particular websites all linked to the Syrian Arab news agency (SANA), which is an official reporting source for the Syrian government. Here is an example:

> #Syria http://t.co/fEsxv6wH news reporter in Idlib: armed terrorist groups hijacked four units of the public order brigade at Bab Alhawa checkpoint #Syria

> مراسل الإخباريه في ادلب: مجموعات إرهابيه مسلحه تختطف عناصر حفظ النظام على باب الهوى الحدودي  #سوريا http://t.co/fEsxv6wH

In contrast, News tweets from regular Arabic and English twitterers included links from a wide range of sources, such as Al-Arabiya, Al-Jazeera, BBC, or Washington Post. Here is an example:

> "Alarabiya.net"published scary pictures of A'zaz massacre in Aleppo http://t.co/zlwkuQ6Q #Al-Arabiya #Syria

> "العربيه.نت" تنشر صورا مرعبه لمجزرة اعزاز في حلب http://t.co/zlwkuQ6Q #العربيه #سوريا

Another major difference crosses the Opinion category. This was the most significant category for Arabic twitterers at about 45% of their tweets. In reading the tweets we see that many Arabic twitterers had strong opinions about what was happening on the ground in Syria and they used Twitter to express themselves. Many expressed shock at how the situation evolved toward a civil war. In addition, due to strong religious sentiments across the Middle East, many were sharing prayers and requesting prayers for the victims of the civil war. Here is an example:

> Oh gracious lord.. Whoever is in your charge shall never be humiliated please extend victory to your army in Syria and show us in the near future the end of Bashar … Lord … No one from the Arabs and non-Arabs tyrannize like him…Lord with your ability let him perish #Syria
>
> ياااااارب ياكريم يامن لا يعز عليك نصر جندك ... أرنا فالقريب العاجل نهاية السفاح بشار ... يارب لم يطغ من العرب و العجم مثله يارب بقدرتك اهلكه #سوريا

The frequency of English tweets that we classified as Opinion (25.8%) was lower than that of Arabic tweets, but still the second largest category. This suggests that English twitterers had similarly strong opinions about the civil war. Lastly, Opinion tweets from the SSB were fewer still at 12.4%.

Another standout difference is in the frequency of spam and phishing content. It appears that in the English-speaking world it is common to use major events to attract attention to sales, product marketing, or phishing. About 25% of the English tweet content related to some form of spam/phishing, attempting to leverage interest in the Syrian civil war to gain attention.

Another major difference between the SSB and the Arabic and English tweets was in the category Other. For the SSB this category comprised 31.8% of the content: tweets still associated with Syria, but not clearly related to the conflict. Since the volume was notably high among the SSB, we decided to unpack this category for the SSB tweets only. We checked the distribution of its secondary codes: International context (59%) was somewhat higher than Local context (41%). The sub-category "Other-Local" encompasses the tweets that contain local news about Syria that are not conflict related, for example:

> #Syria Ministry of Education: 13,500 new jobs within the employment program for youth http://t.co/tLAugGrb #vqk
>
> #سوريا وزارة التربية: 13،500 فرصة عمل جديدة ضمن برنامج توظيف الشباب http://t.co/tLAugGrb #vqk

The link in the tweet discusses the plans for an employment program but does not mention the ongoing conflict in the region. This is one of many similar tweets that include news about meetings happening between government officials that do not mention the ongoing war in Syria. The volume of these tweets show how social botnets can influence the online discussion by covering up tweets related to the main event, in other words, by *smoke screening*.

In contrast, the sub-category "Other-International" encompasses tweets that contain news, facts, and any other topic from outside Syria that is not related to the war. An example of this:

> #Syria BBC… Accused of covering up sexual assaults! http://t.co/pvEVLhOq #pil
>
> قناة البي بي سي متهمة في تغطيه الاعتداء الجنسي ... #سوريا http://t.co/pvEVLhOq #pil

The link in the tweet directed to Syria Steps, a Syrian pro-regime website and one of the three frequently linked to websites by the SSB. The article discusses the news of the English DJ and BBC television presenter Jimmy Savile, who committed sexual abuse, which is not related to Syria. In general, the result of our analysis in the SSB's "Other" category shows that one of the SSB goals may have been to flood the hashtag #Syria with irrelevant topics to distract focus from the civil war, in other words, by *misdirecting*. We will return to more details about smoke screening and misdirection in the Discussion section.

**Influence of SSB**

Our last research question was how the SSB might have influenced the discussion on Twitter regarding the Syrian civil war. We decided to consider the top 100 most retweeted tweets for each week of the dataset. We then considered how many tweets the SSB was able to get into the top 100 and what rank those tweets achieved as a basic measure of the potential influence of the botnet. It is fair to note that the SSB tweets are in Arabic and our analysis here covers the full 35 weeks of botnet dataset, which includes a significant number of English tweets. Thus, the SSB may have had more influence among Arabic Twitter users. Still, given the terms used to collect the dataset our analysis can start to show what it takes for a social botnet to begin to have influence.

Figure 3 illustrates the trend of growing influence of the SSB over the 35 week period. Each week we identified all the tweets in the top 100 most retweeted. We then calculated a botnet influence score based on two metrics: (a) rank influence (the sum of ranked positions of bot retweets in the top 100) and (b) RT influence (a magnitude of retweet cascade represented by the sum of frequencies of bot retweets in the top 100). These graphs reveal a number of interesting characteristics. First, the activity of the botnet is quite "underground" during the first third of its lifespan, apparently having no influence in the top 100 most retweeted content. In the second third it begins to have some influence; but the most significant influence does not occur until the very end, when the total number of bots in the botnet is highest and the frequency of tweeting by each bot is superhumanly high.

We performed a content analysis of all the SSB tweets that made the top 100 for each week using the same content categories developed above. Like the bulk of the human tweets, the top three categories were News, Other, and Opinion, with slight differences in percentage distribution that could simply be a function of the smaller number of tweets that the botnet was able to push up into the top 100 most retweeted tweets each week. That the ratios are similar, suggests the SSB was not discriminating among the tweets that it would attempt to push up through high frequency retweeting.

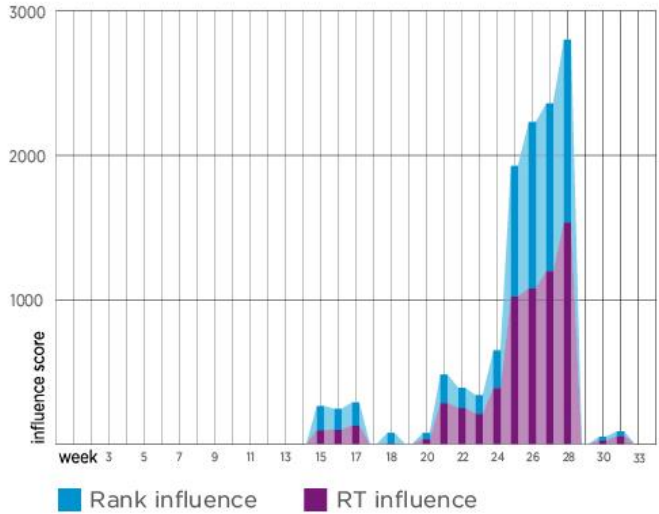

**Figure 3. Influence of the SSB on Twitter RT space.**

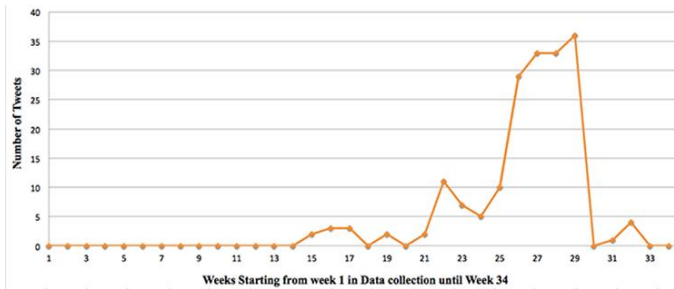

**Figure 4. Number of bots entering the weekly top 100 RT chart.**

For example, in week 29, which was one of the intensely active weeks for the SSB (see Figure 3 and Figure 4), a tweet generated by the bot @LL1 shot up to 4th place in the top 100 most retweeted tweet rankings:

> Media Reporter Hosein Mortada: "Who assassinated General Hassan and a few asked questions: http://t.co/OmagQsV4 #Syria
>
> الاعلامي حسين مرتضى:"قاتل الحسن وأسئلة مطروحة": http://t.co/OmagQsV4 #سوريا

This tweet got retweeted 1041 times by the SSB user accounts only. Another tweet generated by the bot @CB7 mentions the same journalist and reporter Hosein Mortada. This tweet was retweeted 730 times and ranked 12th in week 28 with only the SSB retweeting it:

> The Morning Gate news: the #Lebanese reporter Hussein Morteza told "The Morning Gate": Hezbollah has nothing to do with the killing in #Syria.... http://t.co/wW88DdfL #Syria
>
> بوابة الصباح الاخبارية:الإعلامي ال #لبناني حسين مرتضي لـ"الصباح":لا علاقة لحزب الله و #سوريا بمقتل #سوريا http://t.co/wW88DdfL...

Further analysis shows that the link in each tweet directs to a news article from Al-Sabah (which means "The Morning"), which seems to be an Egyptian news portal.

Additionally, as part of understanding how the SSB influence grew over time, we considered who else was retweeting SSB tweets. We computed the ratio of retweeting between SSB accounts and regular accounts for the top 100 most retweeted tweets. This revealed that 15 tweets out of the 181 SSB tweets that made it in to the top 100 were retweeted by accounts likely in control of a real user (in addition to massive retweeting by the SSB). This is an important finding because it confirms that the SSB's intensive activity got attention from regular twitter users.

We looked closely at all 15 tweets that got retweeted by humans and coded them using our coding scheme.

Similar to the bulk of human tweets, the 15 tweets were within the top two categories News and Opinion. Interestingly, the ones that received the largest amount of retweeting by human accounts were in the Opinion category, which is a category where the SSB was not very active, but humans were. This shows the SSB was able to garner human attention through topics that mirror or mimic humans interest. An example of an Opinion tweet that was mostly retweeted by humans:

> This image made me laugh a lot as it shows Arabs activists participation in #Syria liberating… And in #Bahrain terrorist http://t.co/UyOlqYzl #Syria
>
> أضحكتني هذه الصوره كثيرا لناشطين عرب توضح أن ما يجري في #سوريا ثوره و في #البحرين إرهاب http://t.co/UyOlqYzl

Although the link attached to this tweet is broken, conducting a simple Google search reveals it used to point to a political cartoon. We then computed the ratio of human to SSB retweets for this specific tweet and found that 237 human accounts retweeted it in comparison to 1043 bot accounts, which is an indicator of some human attention.

## DISCUSSION

In this paper we aimed at answering three main questions regarding one specific social bot network on Twitter: to understand how it grows over time, how the content of tweets by the social botnet differ from regular users in the same dataset, and lastly, how the social botnet may have influenced the relevant discussions. Our analysis revealed some interesting findings that sometimes contradict and sometimes support current knowledge on social botnets.

### Mimicking human behaviors

The prior literature on social bots suggests that social bots thrive and succeed by mimicking human behavior on social networks. By mimicking human behavior, bots can go unnoticed and improve their chances of impacting the social graph. The results of our qualitative coding showed that the SSB did not work to mimic human behavior whatsoever. Our analysis shows that over half of the SSB tweets (52.6%) were in the News category, while human tweets (in both English and Arabic) are only about one-third News – a large and very detectable difference. Further in the opposite direction, humans are likely to have a large proportion of their content expressing some personal opinion about an event (45% for our Arabic sample and 25% for our English sample). Based on the data, the SSB does not seem to have much in the way of an opinion about what was happening in Syria, especially when compared to other Arabic language twitterers.

This insight raises two problems with regard to the way the prior work has described and presented ASAs. First is the suggestion that ASAs are becoming sophisticated in construction. Our analysis suggests that this ASA did very little to mimic the activity of a human across both the types of content and the frequency of tweeting activity. While there may be some value in mimicking human activity it may not be completely worth the effort depending on the goals of the ASA operators. We point out that the content most similar to human content was a bit more likely to be retweeted by actual people, but the main question remains.

Secondly, there is a notion that an ASA that mimics human behavior is less detectable. The SSB that we studied was live for over six months. The frequency of tweeting and the connectedness of the retweet social network are what allowed us to identify the bots in the network. While we do not actually know when Twitter detected this specific botnet, we know specifically when Twitter disabled the associated accounts. The challenges for botnet detection are multiple. Determining the motives and legitimacy of the botnet are important. If the botnet is supporting the community, disabling it may be a bad idea. Another challenge is determining how wide a net to cast around the botnet. From our data it seems clear that casting a wide net would include a number of real users who probably would not realize they were part of a botnet simply as a function of their retweeting and following behaviors. Bot and botnet detection is not currently conceptualized as an obvious CSCW topic. But as bots and botnets become participants in larger and more complex social computing systems the relevance to CSCW will grow and what CSCW researchers have to contribute will become more important.

### Misdirection and Smoke Screening

We observed two distinct tactics that the SSB used to point attention away from relevant information (about the Syrian civil war) and influence perception: *misdirection* and *smoke screening*. Misdirection is a technique that magicians use to get the audience to look somewhere else when completing a trick. Similarly, an ASA network that tries to get the audience to attend to other content is using misdirection. In contrast, a smoke screen serves to hide or provide cover for or obscure some type of activity. When the volume of a botnet campaign is high enough with unrelated content, the unrelated content can effectively hide real content.

In the case of our SSB's tweets, many tweets embedded in the Other category were related to foreign news, in particular, political and natural crises happening worldwide. For example, the SSB was benefiting from the political unrest happening in the Kingdom of Bahrain and tweeting news updates on the Bahraini situation:

Pan Ki-moon » calls #Bahraini regime to lift the restrictions on public demonstrations http://t.co/UfMEIPzG #glu #Syria

#سوريا Syria# «كي مون» يدعو النظام #البحريني لرفع القيود عن المظاهرات الشعبية glu# http://t.co/UfMEIPzG

Another example is posting photos of New Jersey and New York damaged by Hurricane Sandy on the east coast of the United States.

Hurricane Sandy Is currently almost 380 miles from New York and is expected to arrive after 5 hours from now http://t.co/mzrexObC #Syria

اعصار_ساندي يبعد حالياً ٣٨٠ ميل تقريباً عن نيويورك ويتوقع وصوله بعد ٥ ساعات من الآن http://t.co/mzrexObC #سوريا

The prevalence of these posts, which were tagged with our collection terms, but seemingly have little to do with the actual events in Syria look to be calculated misdirection, pointing viewers specifically toward other unrelated events. Specifically the SSB was attempting to get people looking for tweets on Syria to pay attention to something major that is happening elsewhere.

This behavior is distinctly different from the way smoke screening is commonly perceived. A few bloggers and news sources such as The Guardian have raised attention to smoke screening on Twitter, describing it as "a cabal of pro-regime accounts, set up for the sole purpose of flooding the #Syria hashtag and overwhelming the pro-revolution narrative" [21].

In the case of the SSB, the remaining tweets embedded in the Other categories were clearly related to events in Syria, just not always related to the civil war. For example:

#Syria 25 short films will be produced by the General Organization for Cinema as part of the 2013 plan to support youth cinema http://t.co/PN2cCEYN#goq # Syria

#سوريا ٢٥ فيلماً قصيراً ستنتجه المؤسسة العامة للسينما ضمن مشروع دعم سينما الشباب للعام 2013 goq# http://t.co/PN2cCEYN #سوريا

This work and the examples we have provided begin to highlight the shades of difference between misdirection and smoke screening as useful techniques to influence social media attention.

Although a full elaboration is out of the scope of this study, it is worth mentioning that misdirection and smoke screening are distinct from astroturfing. The common description of astroturfing is to use a set of ASAs to tweet or retweet a particular point of view to suggest broad consensus around that view. Misdirection and smoke screening rely on other, potentially irrelevant ideas, as opposed to contesting ideas, and tweets or retweets those ideas suggesting that the audience should find those ideas more interesting.

*News reporting*

Throughout the coding stage, the Arabic and English coders were comparing notes and discussing differences between the three samples: SSB, Arabic, and English. As expected, there were clear differences in the News category between English and Arabic tweets, which seem related to differences in language and cultural traditions. However, we were not expecting to see the notable difference between Arabic news reporting and the SSB news reporting as they are both in Arabic. In the latter, the main difference was the frequent use of strong language in the SSB tweets about the opposition, the Free Syrian Army (FSA), and countries that support the FSA, by frequently calling them terrorists and posting torture videos. Here are two examples:

#Syria http://t.co/fEsxv6wH news reporter in Idlib: armed terrorist groups hijacked four units of the public order brigade at Bab Alhawa checkpoint #Syria

Qaeda member Maen Alrez death moment in Homs #Syria #Bahrain #syria #Homs http://t.co/FhFJBwaU #سوريا

*Lifespan model*

One of the popular conceptualizations about social media botnets is that they explode onto the social scene and immediately influence social media discussions. However, the lifespan model of the SSB was somewhat different. The botnet was not built in a day: tracking the SSB for 35 weeks, we observed two distinct stages of its growth. In the earlier stage, from Week 1 to 13, the behavioral patterns appear to be fuzzy and irregular (see Figure 1). It is likely that the SSB operators were experimenting with prototypes, in particular, the 43 short-lived bots whose median lifespan was only two weeks. After Week 14, however, the pattern begins to stabilize. After Week 14 the median lifespan increased to 28 weeks and the median total tweets per week increased to 3449. Interestingly, Week 14 is the same week when the generator bot (@GB1) first started to tweet. We suspect that @GB1 played pivotal role in systematizing the SSB. By looking closer at @GB1's tweet contents, we found a peculiar hashtag, which is a combination of three random letters (e.g., #wqz, #hzf, and so on), attached at the end of each tweet generated by @GB1. We assume that these hashtags may be some kind of identification code for the tweets but do not know the exact use. These random letter hashtags require further investigation.

## CONCLUSION

This study aimed to understand the behaviors and characteristics of social botnets on social computing and social media platforms. We have presented the results of the three analyses we conducted on the dataset of tweets in which the Syrian social botnet was discovered. The results of the three analyses we conducted on the data set aimed at answering three questions: (i) How did the

SSB grew over time, (ii) how the content of the botnet (in Arabic) differs from the content of regular users tweeting in Arabic or English, and lastly, (iii) how the botnet may have influenced the overall Syrian civil war discussion. Our results revealed several new and interesting findings.

As social computing systems grow in sophistication and popularity, bots and collections of bots will take ever more important roles. The collective efforts of individual bots are already critical to supporting Wikipedia [8]. As bots become more social and as they federate in to networks of collective activity, it is essential for designers and researchers to anticipate and design for the interactions between bots and human participants. This work helps set the stage for that by characterizing what is currently happening in the space of social botnets in a specific social media system, Twitter.

Our findings identified a number of differences between what we observed and the way botnets are portrayed in the prior literature. In this paper, we discussed how the SSB in our study was exceptionally long-lived compared to the life span of other reported botnets. This could be due to the fact that the SSB in our study was tweeting in Arabic. This result suggests that there is room for research on the effective detection of social botnets in social media. In addition, from our qualitative coding, it is not clear that the SSB was actually trying to mimic and replicate human behavior. Instead, it seemed mostly interested in flooding Syrian civil war related hashtags with topics that are not war related. That is different from what have been previously discussed in the literature on social bots.

There is an opportunity to continue expanding on this study to answer questions we had during our research. For example, what were the topics the SSB focused on to distract attention from the Syrian civil war? How can the results of this research help increase awareness to this phenomenon and teach SNS users on the dangers of social botnets? In addition, we believe that the results of this work will have a positive impact on the research being conducted on more sophisticated detection software on SNS.